\documentclass[]{article}
\usepackage{amsmath}
\usepackage{amssymb}
\usepackage{epsfig}
\usepackage{subfigure}
\usepackage{umlaut}
\begin{document}
\title{Tree Parity Machine Rekeying Architectures}
\author{Markus Volkmer and Sebastian Wallner\\
Hamburg University of Science and Technology\\
Department of Computer Engineering VI\\
D-21073 Hamburg, Germany\\
\texttt{\{markus.volkmer,wallner\}@tuhh.de}\\
}
\date{}
\maketitle
\begin{abstract}
The necessity to secure the communication between
hardware components in embedded systems becomes increasingly important with regard to
the secrecy of data and particularly its commercial use. 
We suggest a low-cost (\mbox{i.e. small} logic-area) solution for
flexible security levels and short key lifetimes. The basis is an approach for symmetric key exchange using the
synchronisation of Tree Parity Machines. 
Fast successive key generation enables a key exchange within
a few milliseconds, given realistic communication channels with a limited bandwidth.
For demonstration we evaluate characteristics of a standard-cell ASIC
design realisation as IP-core in $0.18\mu$-technology. 
%
\end{abstract}
\section{Introduction}
For embedded systems like handheld devices, smartcards, mobiles or other
wireless communication devices security concepts need to be developed,
in order to keep privacy and still (commercially) exploit the merits of such devices in
widespread and everyday use \cite{A01}. 
This is, for example, of particular interest for the smartcard- or
RFID-industry, where the secrecy of data is directly linked to the commercial prosperity of a product.
Also, the economic importance to secure information technology applications
in the automotive area is becoming eminent along with the protection
of firmware, access control, anti-theft protection, up to scenarios like
the hacking of vital vehicle functions such as an antilock braking
system (see e.g. \cite{ESCAR03}).

Yet, the often relatively small size and severe
power consumption constraints of these devices limit the available
size for additional cryptographic hardware components
\cite{bov97,vedweik97,WGP03}. This holds in particular for sensor networks,
RFID-systems and near field communication devices. 
Secure hardware is thus especially demanded for ubiquitous and pervasive
computing, and the need and research efforts manifest in first conferences on
security in pervasive computing \cite{S03}. 

Hardware-cryptosystems are often based on hard-coded secret keys as
the basic secret. It is good common practice to obey the often
cited {\em Kerckhoffs Principle} \cite{K1883} 
(`no security through obscurity') and not base the security of a crypto-system on the secrecy of the device or algorithm it employs.
The security of a system is thus only as strong as the secrecy of the (fixed) keys. 
But some of the most effective attacks on a crypto-system involve no
ciphertext analysis but instead find flaws in the key-management. Furthermore, insecure bus communication as reported in \cite{H02} (regarding
the video game console market), allows attacks still above the chip level by sniffing internal buses.
In embedded system environments, functions are being realized
(at least partly) in hardware and often lack online system access. The changing of a fixed
key, as any other security update, is difficult or even impossible -- i.e. too expensive.

The exchange of a common secret key over a public channel is dominated by
methods based on number theory since the invention of the
Diffie-Hellman key exchange protocol in 1976 \cite{DH76}. 
Computational security is based on the difficulty of the discrete
logarithm problem in \mbox{{\em El Gamal} \cite{ElG85}}, which is considered as
difficult as the factorisation problem of a product of long prime
numbers as in \mbox{{\em RSA} \cite{RSA78}}. Such asymmetric algorithms need to perform a lot of
computational intensive arithmetics on typically limited embedded
microcontrollers. In a particular GSM mobile phone, for example, two
algorithms are combined to meet performance requirements: an
asymmetrical algorithm with a 1024 bit key for key exchange and a
symmetrical algorithm using only 128 bit for the key and voice encryption \cite{TopSec}.  
This also demonstrates the often necessary tradeoff between the level
of security and the available resources.
 
The state-of-the-art, regarding applications in embedded systems, is represented by {\em Elliptic Curve Cryptography} and the generalisation to
{\em Hyper-Elliptic Curves} (see e.g. \cite{PWP03}). Without a reduction of
the security, these representations allow to reduce the size of the numbers to calculate with. Yet, more complex
expressions need to be calculated. 
After all, a (frequent) key exchange is often of prohibitive cost, especially in the often changing topology of
pervasive or ad-hoc networks.

In this paper we present a small hardware solution for secure
data exchange with flexible security levels and short key lifetimes. It is based on a fast successive 
key generation and exchange process. We use a hardware-friendly
algorithm for secure symmetric key exchange by synchronisation of socalled {\em Tree Parity
Machines} \cite{KKK02}.
We define architectures, using this key exchange concept, that allow fast successive key generation 
and exchange. The key exchange ranges within milliseconds for
realistic channels and can be performed in parallel (or multiplexed) to encryption and the encrypted
communication process. 
Additionally, we provide the architectures with a flexible rekeying
functionality to enable full exploitation of the achievable exchange
rates. This particularly increases the cost for a successful immediate (online)
attack, as opposed to a subsequent (offline) analysis on recorded information.  
Our focus is on secure data exchange between hardware components in
embedded systems like RAM, FLASH-type ROM, (co-)processors and on
bus-communication in general. Environments in which security can also be of
moderate concern are also considered. 
 
In the following, we introduce the neural network structure and a
learning algorithm (section \ref{nnalg}), also in order to already point out advantageous properties for a hardware realization. The
synchronization effect leading to the key exchange property is
explained. Algorithmic security implications on the realization of our
architecture are described in section \ref{decision}.
Section \ref{architec} comprises the architectural design of the
proposed hardware component with its rekeying functionality. Here, we also refer to design decisions
prepared in the previous section. 
In section \ref{resul}, we present results from an FPGA and an ASIC
implementation on silicon area, possible clock and
key exchange rates (throughput).
We conclude the paper in section \ref{conc} with a short summary and an outlook on possible further
extensions also referring to current research activities.  
\section{Tree Parity Machines for Key Exchange}\label{nnalg}
\begin{figure}[ht!]
\begin{center}
\subfigure[]{\epsfig{file=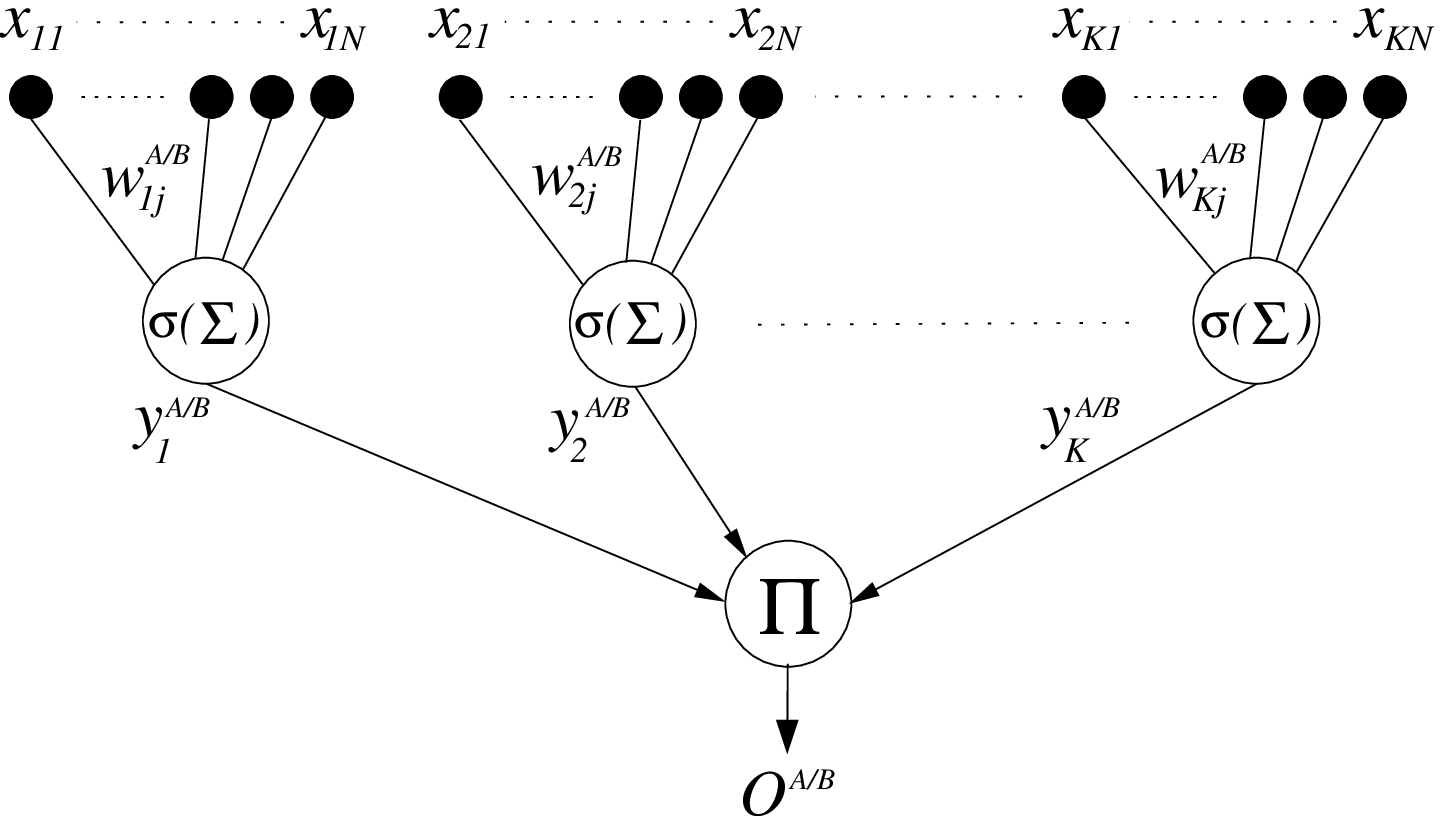,height=4.0cm}}\hspace{1cm}
\subfigure[]{\epsfig{file=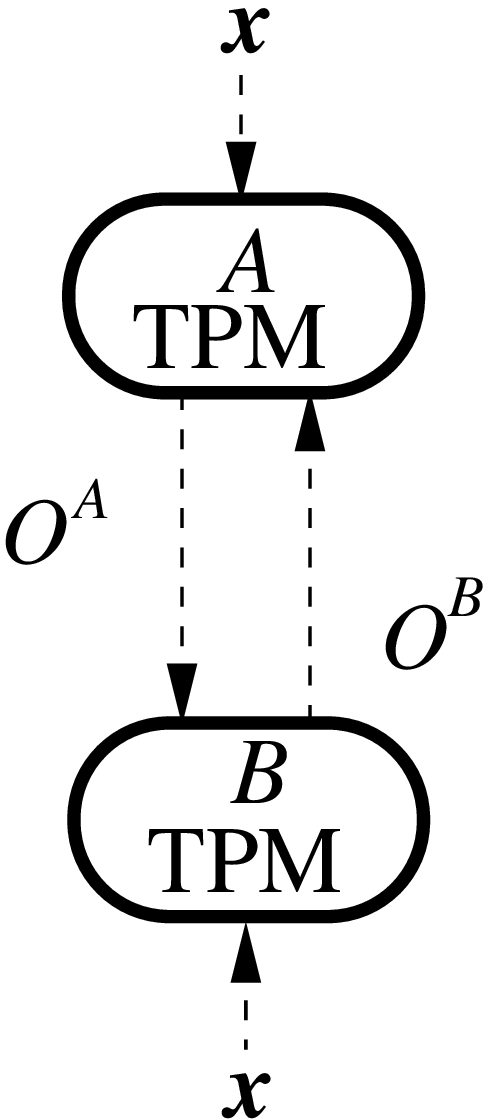,height=3.6cm}}
\caption{(a) The tree parity machine (TPM) generates a single output -- the parity of the outputs of the hidden units. (b) For mutual
learning, outputs on commonly given inputs are exchanged between the
two parties $A$ and $B$.}
\label{PM_fig}
\end{center}
\end{figure}

In \cite{KKK02}, Kinzel et.~al. proposed a symmetric key exchange method based on the fast synchronisation of two
identically structured Tree Parity Machines (TPMs). The particular tree structure 
has non-overlapping binary inputs, discrete weights and a single
binary output as depicted in \mbox{Fig. \ref{PM_fig}a}.
Studying interacting neural networks in general (cf. \cite{MKK00,KK02,K03}), the authors focused
the phenomenom of fast synchronization by mutual learning TPMs and its potential
for a cryptographic approach, not involving large numbers and methods from number theory.
Their exchange protocol is realized implicitly by a mutual adaptation process
between the two parties $A$ and $B$, not involving large numbers and methods from number theory. 
\subsection{Structure of a Tree Parity Machine}\label{tpmnn}
In the following, we describe the implemented parallel-weights version using
hebbian learning (cf. \cite{KKK02
}). Weights are identical in both TPMs after synchronisation.
The anti-parallel-weights version, using anti-hebbian learning and
leading to inverted weights at the other party, is omitted for brevity.
The notation $A/B$ denotes equivalent operations for the parties $A$ and $B$. A single $A$ or $B$ denotes an operation which is
specific to one of the parties.

The TPM (see \mbox{Fig. \ref{PM_fig}a}) consists of $K$ hidden units ($1\leq k\leq K$)
in a single hidden-layer with non-overlapping inputs (the
tree structure) and a single unit in the output-layer.

Each hidden unit receives different $N$ inputs ($1\leq j\leq N$), leading
to an input field of size $K\cdot N$. The vector-components are random variables with zero mean and unit variance.
The output $O^{\scriptscriptstyle{A/B}}(t)\in\{-1,1\}$, given
bounded weights \mbox{$w_{kj}^{\scriptscriptstyle{A/B}}(t)\in[-L,L]\subseteq \mathbb{Z}$}
(from input unit $j$ to hidden unit $k$) and common random inputs $x_{kj}(t)\in\{-1,1\}$, is calculated by a parity function of the
signs of summations:
\begin{small}\begin{equation}\label{out}
O^{\scriptscriptstyle{A/B}}(t)=\prod_{k=1}^{K}y^{\scriptscriptstyle{A/B}}_k(t)=\prod_{k=1}^{K}\sigma(\alpha^{\scriptscriptstyle{A/B}}_k(t))=\prod_{k=1}^{K}\sigma\!\left(\sum_{j=1}^{N}w_{kj}^{\scriptscriptstyle{A/B}}(t)\
x_{kj}(t)\right)\ .
\end{equation}\end{small}
The common random inputs can also be kept secret between the parties,
yielding authentication (see Section \ref{bplearn}).  
$\sigma$ is a party-specific modified sign-function, that defines an
agreement between the two parties on an opposite sign in case of a sum $\alpha^{\scriptscriptstyle{A/B}}_k(t)\in\mathbb{Z}$ of zero:
\begin{small}\begin{equation}\label{branch}
\sigma(\alpha^{\scriptscriptstyle{A/B}}_k(t)):=\begin{cases}\ \ 1\quad , \alpha^{\scriptscriptstyle{A/B}}_k(t)>0\ \vee\ \alpha^{\scriptscriptstyle{A}}_k(t)=0\\
-1\quad , \alpha^{\scriptscriptstyle{A/B}}_k(t)<0\ \vee\
\alpha^{\scriptscriptstyle{B}}_k(t)=0\ .
\end{cases}
\end{equation}\end{small}

From the communicated output, the outputs of the hidden units cannot
be uniquely determined. There are multiple combinations for a
signed or unsigned output, depending on the number of hidden units $K$.

\subsection{Key Exchange by Mutual Learning and Synchronisation}\label{bplearn}
The so-called {\em bit package} variant was chosen for implementation \mbox{(cf. \cite{KKK02})}.
Due to an reduction of (physical) output exchanges by an order of
magnitude, it is advantageous for practical communication channels
with a certain protocol overhead.

Parties $A$ and $B$ start with an individual randomly generated initial
weight vector $w_{kj}^{\scriptscriptstyle{A/B}}(t_{\scriptscriptstyle{0}})$ -- their
secret. After a set of $b>1$ presented inputs, where $b$ denotes the size of the
bit package, the corresponding $b$ TPM outputs (bits)
$O^{\scriptscriptstyle{A/B}}(t)$ are exchanged over the public
channel in one package (see \mbox{Fig. \ref{PM_fig}b}). The $b$ sequences of hidden states
$y^{\scriptscriptstyle{A/B}}_k(t)\in\{-1,1\}$ are stored for the subsequent learning process.

A hebbian learning rule is applied to adapt the weights, using
the $b$ outputs and $b$ sequences of hidden states. They are
changed only on an agreement on the parties' outputs. Furthermore, only weights of 
those hidden units are changed, that agree with this output: 
\begin{small}\begin{equation}\label{learn}
w_{kj}^{\scriptscriptstyle{A/B}}(t):=\begin{cases}w_{kj}^{\scriptscriptstyle{A/B}}(t-1)+O^{\scriptscriptstyle{A/B}}(t)\
x_{kj}(t)&,O^{\scriptscriptstyle{A}}(t)=O^{\scriptscriptstyle{B}}(t)\
\wedge\\&\ O^{\scriptscriptstyle{A/B}}(t)\ y^{\scriptscriptstyle{A/B}}_k(t)>0\\
w_{kj}^{\scriptscriptstyle{A/B}}(t-1)&,\textrm{otherwise.}\end{cases}
\end{equation}\end{small}

Updated weights are bound to stay in the maximum range
\mbox{$[-L,L]\subseteq \mathbb{Z}$} by reflection onto the boundary values
\begin{small}\begin{equation}\label{bound}
w_{kj}^{\scriptscriptstyle{A/B}}(t):=\begin{cases}\textrm{sign}\!\left(w_{kj}^{\scriptscriptstyle{A/B}}(t)\right)L&, |w_{kj}^{\scriptscriptstyle{A/B}}(t)|>L\\
w_{kj}^{\scriptscriptstyle{A/B}}(t)&,\textrm{otherwise.}\end{cases}
\end{equation}\end{small}

In iterating the above procedure, each component of the weight vectors performs a random walk
with reflecting boundaries. This implies a trajectory in a weight space of \mbox{$(2L+1)^{\scriptscriptstyle{K\!N}}$} points. 
Two corresponding components in $w_{kj}^{\scriptscriptstyle{A}}(t)$ and
$w_{kj}^{\scriptscriptstyle{B}}(t)$ receive the same random
component of the common input vector $x_{kj}(t)$. After each bounding
operation (Eq. \ref{bound}), the distance between the components is
successively reduced to zero. 
Synchrony is achieved when both parties have learned to produce
each others outputs. They remain synchronised (see learning rule \mbox{Eq. (\ref{learn}})) and
continue to produce the same outputs on every commonly given input.
This effect in particular leads to common weight-vectors in both TPMs
in each of the following iterations. These weights have never been communicated
between the two parties and can be used as a common time-dependent key for
encryption and decryption respectively. Such secret key agreement based on interaction over a public
insecure channel is also discussed under information theoretic aspects
by Maurer \cite{Maurer93a}, especially with regard to unconditional security. Furthermore, synchrony is
achieved only for {\em common} inputs. Thus, keeping the common inputs
secret between $A$ and $B$ can be used to have an authenticated key
exchange. There are $2^{KN}-1$ possible inputs in each iteration,
yielding as many possible initialisations for a pseudo random
number generator. 
Shamir et al. conferred to
such a synchronization over multiple rounds as {\em a
gradual type of Diffie-Hellman} key exchange
\cite{KMS02asia}, because Diffie-Hellman has a single round that transmits several
bits. 
Obviously, a test for synchrony cannot practically be defined by checking whether weights in both
nets have become identical. One therefore tests on successive equal outputs in a
sufficiently large number of iterations $t_{\scriptscriptstyle{min}}$, such that equal outputs by
chance are excluded:
\begin{small}\begin{equation}\label{synccrit}
\forall t\in[t',\cdots,t'+t_{min}]:\
O^{\scriptscriptstyle{A}}(t)=O^{\scriptscriptstyle{B}}(t)\ .
\end{equation}\end{small}
The number of outputs (bits) required to achieve synchronisation is lower than
the size of the key \cite{MPKK02b}. Synchronisation time is finite for discrete weights. It is almost
independent on $N$ and scales with $\ln N$ for very large $N$. 
Furthermore, it is proportional to $L^2$ \cite{MPKK02b}. 
Our investigations/experiments confirmed that the average synchronisation time
is distributed and peaked around 400 for the parameters given in
\cite{KKK02}.
\section{Security and Rekeying Functionality}\label{decision}
The symmetric key-exchange protocol can generate
long keys by fast calculations and building the secure channel is of
linear complexity. It scales with the size $K\cdot N$ of the TPM structure \cite{KKK02}, which defines the size
\mbox{$K\cdot N\cdot L$} of the key.
In order to still allow comparisons with the literature we chose $L=4$
for our implementation.
The time to
synchronise roughly doubles in comparison to $L=3$, while for the attacker the same time increases by
orders of magnitude (see \cite{RKKK02a}).

The security of the key exchange manifests in algorithm-specific
properties and can be fully exploited by appropriate hardware design.  
Next to other general (algorithmic) aspects on the security of the
exchange method as described in \cite{KKK02,KMS02asia}
, the tracking of the weights is hard in comparison to synchronisation -- practically
even harder when implemented in hardware.

The key exchange protocol has been attacked by several eavesdropping approaches,
which always require full knowledge of the TPM structure in use. We
will describe their basic properties in order to clarify the security
aspect. Due to the nature of the key exchange and its attacks, only probabilistic
definitions of a `successful attack' can be provided. One can distinguish between two classes of
attacks. The first class comprises attacks, that can be defeated by
appropriately increasing the parameter $L$. The consequence is, that
the learning time of an attacker is significantly longer than the
synchronisation time. 
The security increases proportional to $L^2$ while the probability of a successful attack decreases exponentially with
\mbox{$L$ \cite{MPKK02b}}. Among these so defeatable attacks, which try to synchronise faster than the two parties
\cite{KK02NNconf}, are the {\em Naive Attack}, that uses a single or an ensemble of several identically
structured TPMs. The {\em Genetic Attack} even comprises a population of
thousands of TPMs, whose internal representations are optimised by a
genetic algorithm \cite{KMS02asia}. A successful attack is
defined here as synchronising faster than the parties $A$ and $B$ and could be realized for
$K=2$ in $50$\% of all cases. But, already for $L=3$, this
attack has shown to be less effective than the {\em Flipping
Attack}. The complexity of such attacks (especially in hardware),
with hundreds or thousands of TPMs plus an additional (genetic) algorithm, is obviously high.
The Flipping Attack defines a successful attack as having $98$\%
overlap with the weights of one party, when parties $A$ and $B$ are
already synchronous. For $L>2$,  an ensemble of $10000$ Flipping Attackers was found
less effective than a single attacker, which revokes its practical use
\cite{KK02NNconf}.

All of the previously sketched attacks can be made arbitrarily costly
and thus practically defeated by increasing L, which significantly decreases the
probability of a successful attack. The approach thus remains computationally secure for sufficiently large L \cite{RKKK02a,SKMKK03}.  

The only attack, which does not seem to be affected by an increase of
$L$ (but still by an increase of K) is the socalled {\em Majority Flipping Attack}. It uses a hundred of coordinated and
communicating TPMs \cite{SKMKK03}. Yet, the given definition of a successful
attack is problematic: When A and B have fully synchronised, the attacker has 98\% average overlap (i.e. a fraction) with the weights, in 50\% of all
cases. For 99\% average overlap, the probability reduces to 25\%. This indicates that the difficulty lies in achieving the last percents. The authors chose
this definition, because of the strong fluctuations they observed in
the success probability. But the definition of overlap is an average
overlap over all hidden units. Thus an attacker does not know, which
of the $K\cdot N$ components of the weights (the key) are correct in a real attack scenario. In currently used symmetric encryption algorithms, the
flipping of a single bit only already leads to a complete failure in
decryption. Thus practically, one still has to perform a subsequent
brute-force attack, which would then only be successful in 50\% of all cases.
Keep in mind, that $A$ and $B$ already have one key (while the
attacker has 98\% with a probability of 0.5) and start encryption
and data transmission. The attacker needs to perform his brute force attack plus the attack on the encrypted data in parallel.  
Furthermore, the rekeying principle and the achievable short key
lifetimes (cf. Section \ref{rekeyprinc}) aim at an online usage of the
exchanged keys for secure transmission.

All formulated attacks can hardly be performed online. Only an
offline attack on the previously recorded exchanged information seems realistic.
Last but not least, note that all of the existing attacks are based on
knowing the common inputs and thus refer to a non-authenticated key
exchange, in which man-in-the-middle attacks are possible as well. 
\subsection{Feasible Immediate Rekeying}\label{rekeyprinc}
We propose to minimise the key lifetime as much as possible employing
{\em immediate rekeying}, that allows to exploit the speed of key exchange
and features of our hardware component. Such a rekeying process normally is to be
avoided due to the computational cost of a new key
exchange. Strategies are developed to increase the key lifetime
without affecting the security (see e.g. \cite{AB00}). 
Yet, using the TPM principle allows for efficient rekeying in the
$k\!H\!z$-range (see also \mbox{Section \ref{resul}}).

Next to several other propositions (cf. \cite{KKK02}) concerning the en-/decryption, 
one particular proposition (cf. \cite{KK02
}) 
is to take each (common) weight vector after synchronisation for
en-/decryption. On the one hand, a new potential key is present in
each step, which can then be used block-wise. On the other hand, an opponent then also has the
chance to synchronize using 	the ongoing communication (cf. \mbox{Section
\ref{bplearn}}) and get a key.

We suggest to permanently generate (i.e. synchronise)
new keys in parallel or multiplexed to the encryption-transmission-decryption of
data, using the most recent key exchanged. In this case, the key is only used to encrypt a certain
small subset of the plaintext. As soon as a new key has been exchanged, it is used for encryption. 
This especially allows to realize short key lifetimes, enabling a
certain security level by many smaller keys instead of one large key. 

Consequently, in our hardware design, we allow an external unit to demand a key exchange
service. Our TPMRAs will continuously synchronise new keys, as long as
data needs to be exchanged. Once a crypt-unit uses the first key,
synchronisation is triggered again to always provide a next key.
In this way, the related hardware resources are consequently used and
keys are exchanged at a maximum rate subject to hardware
constraints, average synchronisation time and available channel bandwidth. Furthermore, it allows
to implement services like periodic or even adaptive rekeying.
Security is thus increased in our hardware implementation through
feasible immediate rekeying, the mere speed of key
exchange and the achievable short key lifetimes. 
\section{Tree Parity Machine Rekeying Architectures}\label{architec}
It is important to note that, with respect to a hardware implementation, only signs and
bounded integers are processed within the algorithm. The result of the outer product in \mbox{Eq.
(\ref{out})} can be realized without multiplication. The product within the sum is
only changing the sign of the weight. Thus, the most complex structure to be
implemented is an adder. The complexity of such a unit is thus even less than the complexity of a
linear filter, which requires a full multiply-accumulate structure. Yet, the inherent parallelism can be exploited here as well. The branches in \mbox{Eq. (\ref{branch})} are only
based on a test for the sign or a test on equality to zero, also easily
done in hardware. 

Furthermore, only sign-operations and additions are present in the
learning rule (\mbox{Eq. (\ref{learn})}, well suited for a hardware implementation. The bit
package exchange can either be realized serially or via a parallel bus,
depending on the users requirements and the intended application. The
amount of registers needed for storage increases in the bit package
variant, finally imposing a tradeoff area vs. speed.

Equal (pseudo-)random inputs are realized by equally initialised Linear Feedback
Shift Registers (LFSR) or a Cyclic Redundancy Code (CRC). 
Different (secret) initial weights can either be fixed
(device-specific), or they can be provided by an additional
application-specific device or by a thermal noise device. The synchronisation
criterion (\mbox{Eq. (\ref{synccrit})}) basically comprises a counter.

The proposed Tree Parity Machine Rekeying Architectures (TPMRAs) are functionally separated into two main
structures. One structure essentially comprises the Key Handshake and Bit Package
Control. The other structure contains the TPM Unit and its control state machine.
\subsection{Key Handshake and Bit Package Control}
\begin{figure}[htb]
\begin{center}
\includegraphics[scale=0.45]{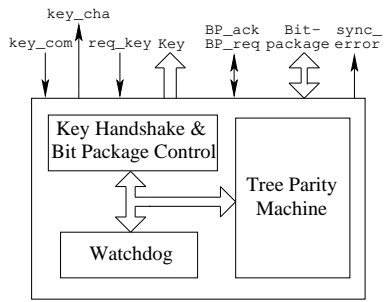}
\end{center}
\caption{Basic diagram of the Tree Parity Machine Rekeying Architectures.\label{architecture}}
\end{figure}

As described in \mbox{Section \ref{bplearn}}, we implemented the {\em
bit package} generalisation of the protocol (cf. \cite{KKK02}). The
overall structure of the TPMRAs is shown in \mbox{Fig. \ref{architecture}}. It consists of three functional blocks:
a Key Handshake and Bit Package Control, the TPM unit and a Watchdog timer. 

The Watchdog timer supervises the number of interactions needed
for a key-exchange between two parties (\mbox{Eq. (\ref{synccrit})}). If there is no synchronisation
within a specific time (remember that the synchronisation time is
distributed), a signal ({\tt sync\_error}) indicates a synchronisation
error. It is programmable for variable average synchronisation
times subject to the chosen TPM structure.

The Key Handshake and Bit Package Control handles the key transmission with an encryption
unit and the bit package exchange process with the other party. It accomplishes the bit packaging by partitioning the parity bits from
the TPM unit in tighter bit slices. Due to different computation cycles between two key exchange parties,
the rekeying procedure employs a key request ({\tt req\_key}), a key changed
({\tt key\_cha}) and a key commit ({\tt key\_com}) handshake protocol (see \mbox{Fig. \ref{architecture}}). A key is
handed over via the internal bus ({\tt Key}) to an encryption unit when the synchronisation process is finished. For our application
domain in embedded system environments, we choose a fixed bit package
length of 32 bit for physically parallel exchange and synchronisation
over a 32 bit wide bus ({\tt Bit Package}). The bit package exchange process uses a simple
request/acknowledge handshake protocol ({\tt BP\_ack}, {\tt BP\_req}). 

\subsection{Tree Parity Machine Unit}
\begin{figure}[htb]
\begin{center}
\includegraphics[scale=0.45]{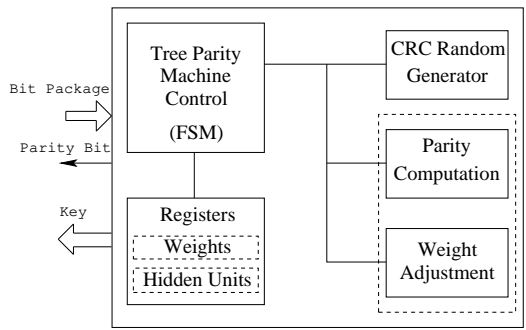}
\end{center}
\caption{Internal structure of the Tree Parity Machine Unit.\label{tpmunit}}
\end{figure}

The TPM unit comprises the logic for the TPM structure, such as the logic for calculating the parity bits as explained in
\mbox{Section \ref{nnalg}}. It consists of the TPM control, a Cyclic
Redundancy Code (CRC) generator, a Parity Computation unit and a Weight Adjustment unit. A register bank holds the
data for the hidden unit and the weights of the
network as shown in \mbox{Fig. \ref{tpmunit}}. 

The TPM control is realized as simple finite state
machine (FSM) which executes the initialisation of the TPM and the learning
process with the bit package from the other party. The Parity Computation unit
calculates the summation and the parity bit (\mbox{Eq. (\ref{out})} and
\ref{branch}). The weight adjustment unit accomplishes the learning
rule (\mbox{Eq. (\ref{learn})} and (\ref{bound})). 

The CRC random generator generates the pseudo random bits for the
inputs of the TPM. It is initialised by a vector which is equal for
both parties. For the purpose of authentication, the initial
value would have to be kept secret. 

\section{Implementation and Performance}\label{resul}
We designed and simulated parameterisable, serial and semi-parallel TPM
Rekeying Architectures, using VHDL to implement an FPGA- and an ASIC-realisation. For both architectures we appoint
the integer range $L$ to $4$, as explained in \mbox{Section \ref{decision}}. 
In the serial architecture, the synaptic summation is performed by Time
Devision Multiple Access (TDMA) of an $L$-bit adder, while the semi-parallel form uses
TDMA of six $L$-bit adders in parallel. The details of the
TPMRA implementations (key length $K\cdot N\cdot L$, serial or semi/fully-parallel
realisation) must be chosen with respect to the target environment, including the used parameters, the timing, the available channel capacity
and the available chip-area, of course. 

We realized a two party prototype system based on two \mbox{{\em XCV300E-8}}
{\em (Virtex E)} FPGAs from {\em Xilinx} in order to investigate and demonstrate the functionality of our
architectures.  
Standard cell ASIC prototype realisations were built to verify the suitability
of the TPMRAs for typical embedded system components. We chose $K=3$
and varied $N$ up to $49$ for a resulting key size of $3\cdot 49\cdot 4=588$ bit. This choice for $N$ already allows a remarkable
key length and still keeps the average synchronisation time low (cf. \cite{KKK02}). The underlying process
was a $0.18\mu$ six-layer CMOS process with $1.8V$ supply voltage
based on the {\em UMC library}. 
The design was synthesised using the
{\em Synopsys Design-Compiler} 
and was mapped using {\em Cadence Silicon Ensemble}.

\begin{figure}[ht!]
\begin{center}
\subfigure[{\mbox{Area [$mm^2$]} vs. key \mbox{length [$bit$]}}]{\includegraphics[width=5.8cm]{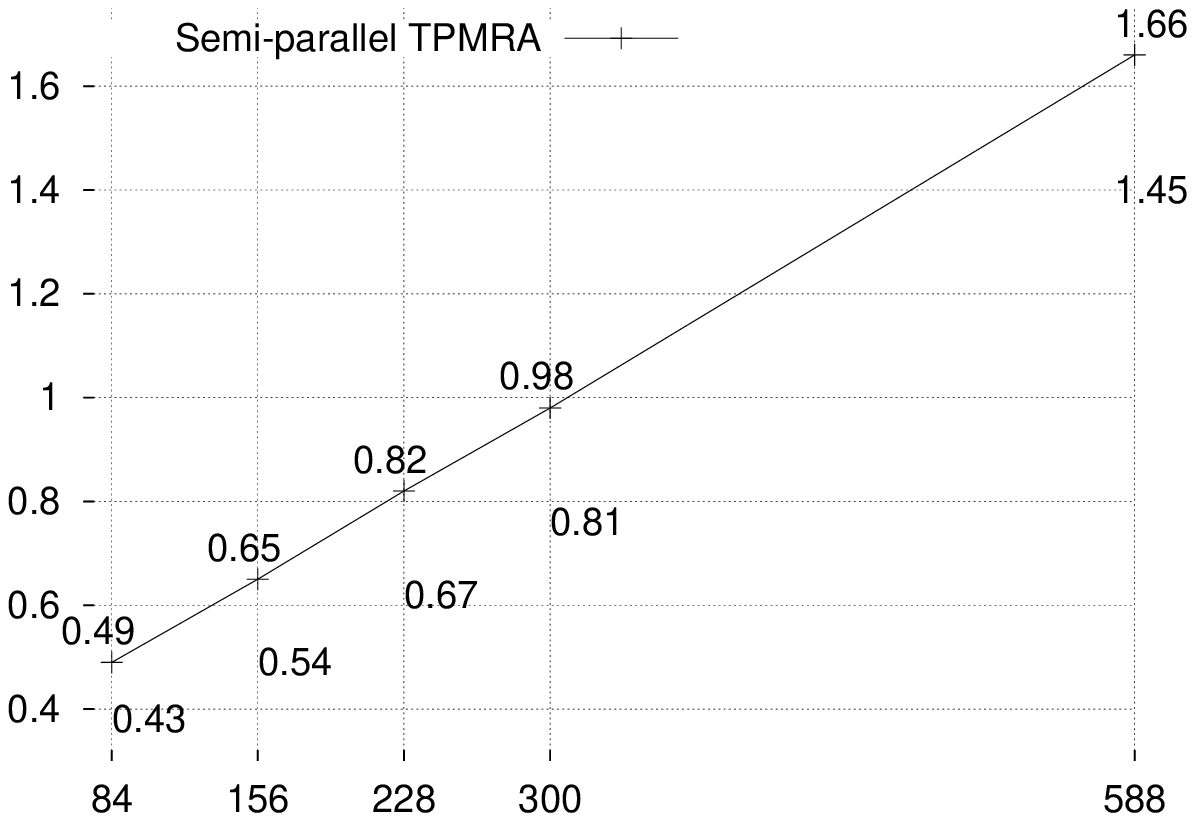}}\hspace{0.01cm}
\subfigure[{\mbox{Speed [$M\!H\!z$]} vs. key \mbox{length [$bit$]}}]{\includegraphics[width=5.8cm]{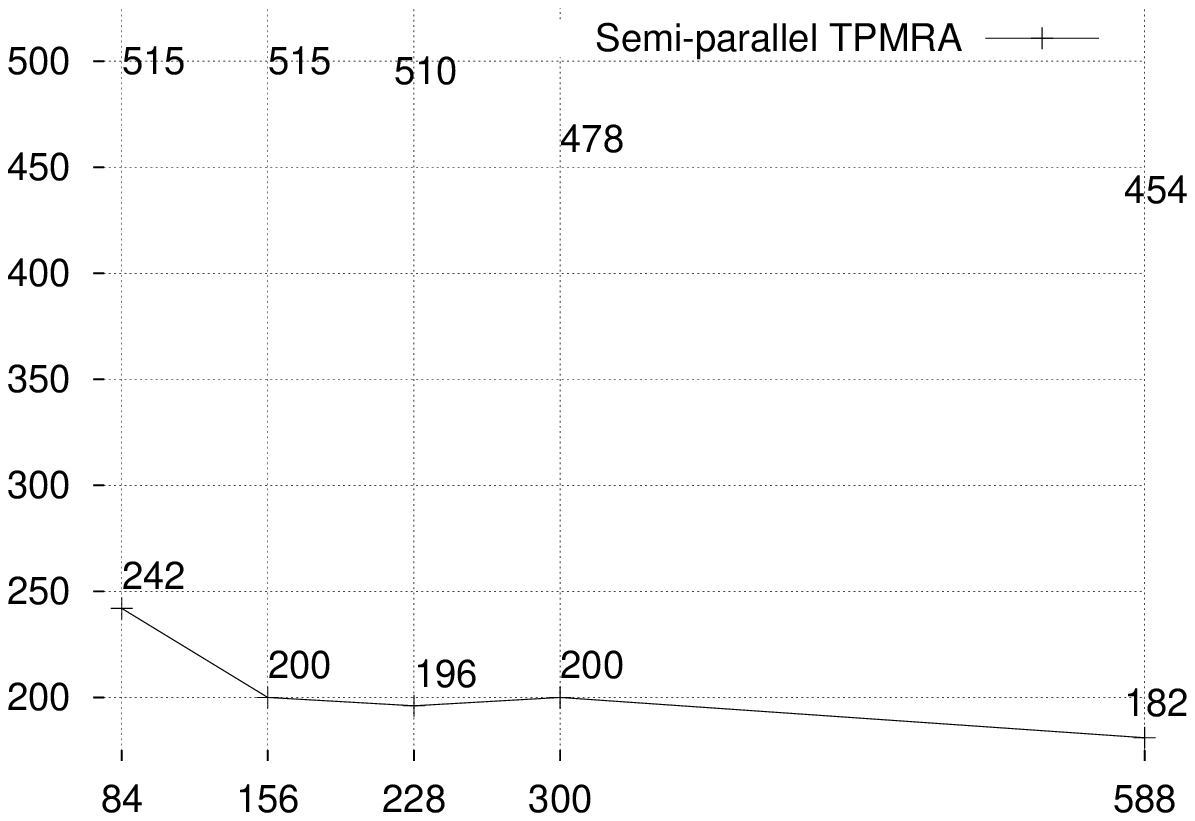}}\\
\subfigure[{Average key exchange \mbox{rate [$H\!z$]} vs. key \mbox{length [$bit$]}
(idealised infinite channel bandwidth)}]{\includegraphics[width=5.8cm]{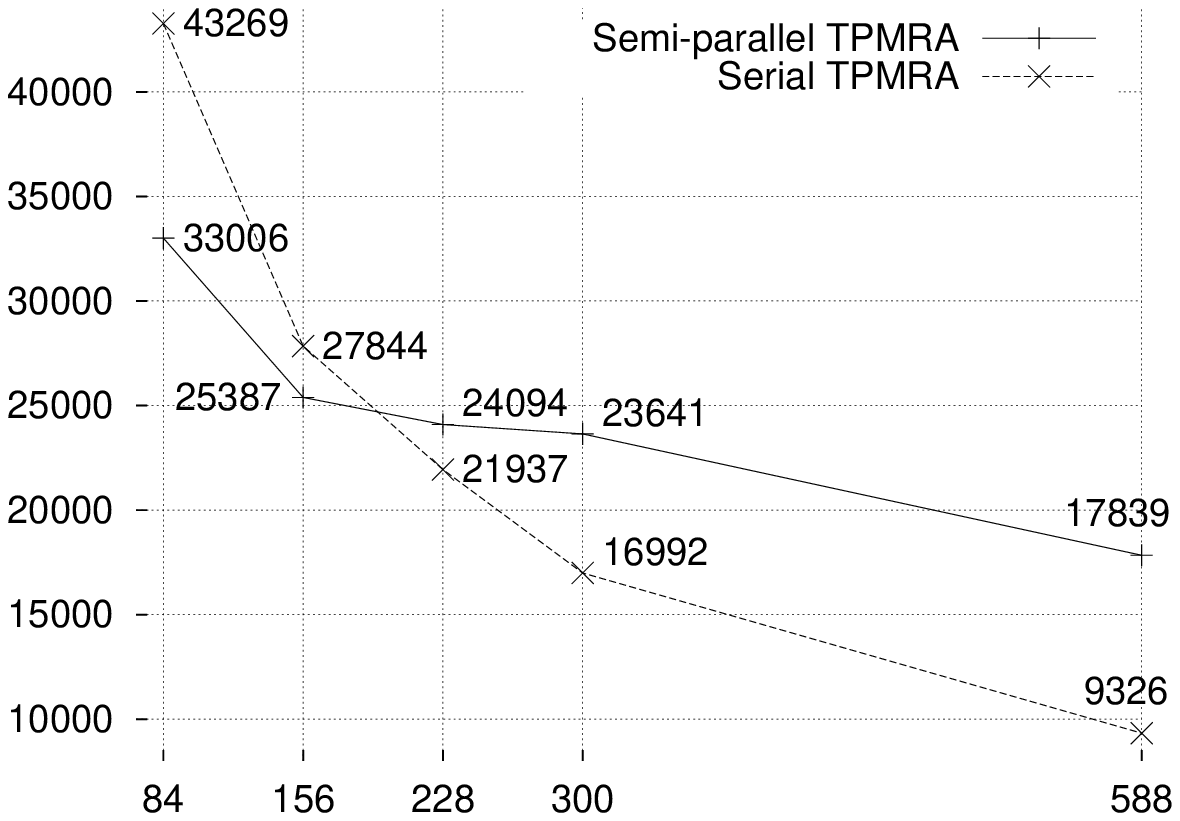}}\hspace{0.01cm}
\subfigure[{Average key exchange \mbox{rate [$H\!z$]} vs. the channel bandwidths [$kbps$] of $I^2C$, $C\!AN$ and $PCI$ (burst mode)}]{\includegraphics[width=5.8cm]{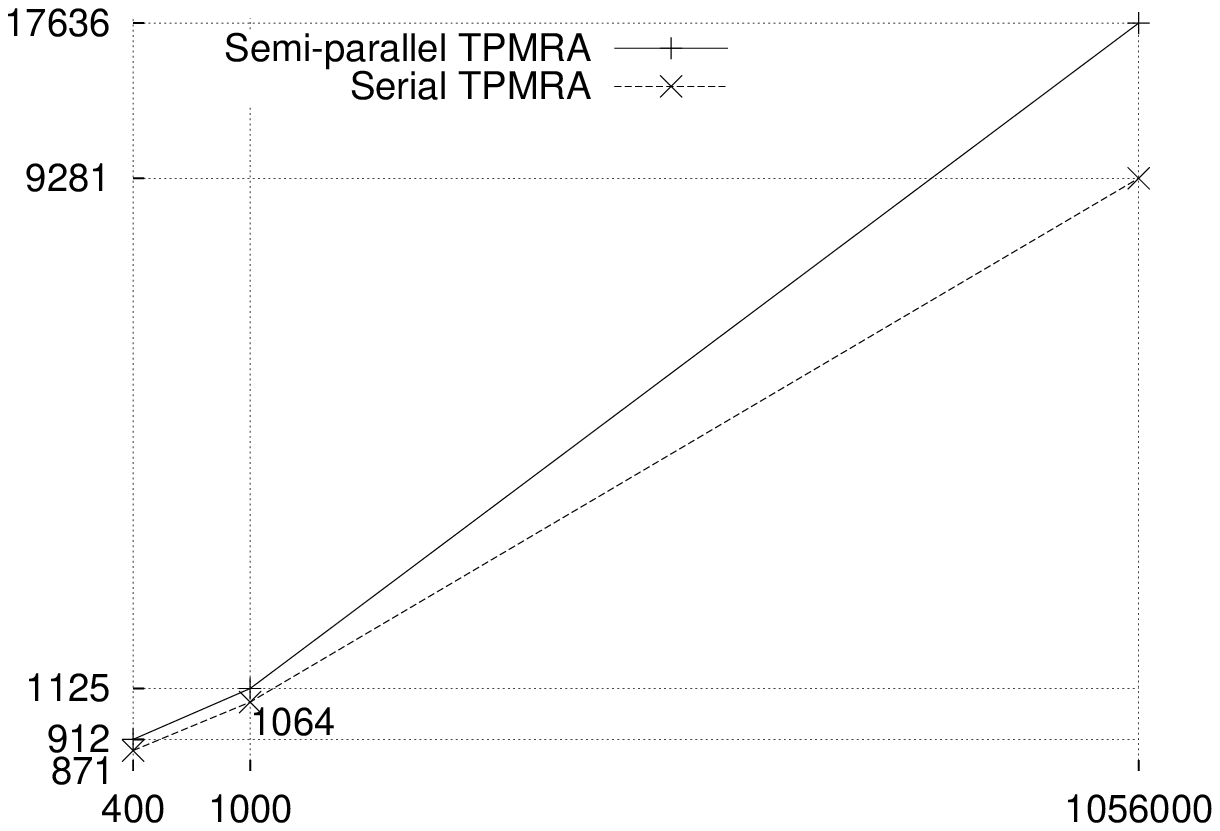}}
\end{center}
\caption{Post-synthesis results for chip-area (logic) (a) and
achievable clock-frequency (b) vs. key length. Average key exchange
rate (avg. synchronisation time of 400 iterations) vs. key length is plotted in (c). A practically finite
channel capacity is neglected here. Plot (d) is log-scaled and shows average key
exchange rates for a 588 bit key and a selection of typical channels with their capacities. All data refers to a
UMC 0.18 micron six-layer standard cell process.\label{areaspeed}}
\end{figure}
The area (\mbox{Fig. \ref{areaspeed}a}) of the TPMRA realisations scale approximately
linear (around one square-millimetre) due to the linear complexity of the adders. The serial TPM
realisation consumes less area (i.e. less hardware resources). Note, that most of the area is consumed by the bit
packaging, because of the necessary storage of the inputs for the learning
(cf. \mbox{Section \ref{bplearn}}). 

Obviously, the achievable
clock speed (\mbox{Fig. \ref{areaspeed}b}) in the serial variant is significantly higher than in the
semi-parallel version. This is due to the necessity of a longer clock
tree for the additional registers to store partial results. 

Additionally, we established the throughput (i.e. keys per second) subject to the average synchronisation
time of 400 iterations for different key lengths in
\mbox{Fig. \ref{areaspeed}c}. We assumed the maximally achievable clock frequency
with regard to each key length, which can be achieved by Digital Phase Lock Loop (DPLL), regardless of the systems clock frequency. 
Furthermore, we appointed the average synchronisation time of 400
iterations for all key lengths, although it is really always less than the size
of the key (a worst-case scenario). This data refers to an idealised infinite
channel bandwidth, neglecting the transmission delay.
For key lengths smaller than approximately 180 bit, the serial TPMRA
has a higher throughput (in the range of $2.5\cdot 10^4$ to more than $4\cdot 10^4$ keys
per second) due to the higher clock frequency
(\mbox{Fig. \ref{areaspeed}b}). Beyond this point, the semi-parallel version
achieves a higher throughput, exploiting the parallel computation.   

Figure \ref{areaspeed}d shows the same information, but for three real
communication channels and their bandwidths, given a key length of 588 bit. The
chosen log-scale allows to see the small difference regarding the
throughput (up to around 1000 keys per second) for an $I^2C$ and $C\!AN$-bus. Only for buses of higher
bandwidth such as the $PCI$-bus, the two architectures show a significantly different
throughput (reaching the $k\!H\!z$-range). In the case of an 32 bit $PCI$-bus in burst mode, the theoretical maximum throughput (as in
\mbox{Fig. \ref{areaspeed}c}) can be achieved. We also considered other bus systems (e.g. packet based
systems like WLAN). The results are similar, due to their small bandwidth
in comparison to the $PCI$-bus. Obviously, the bottleneck is the
underlying communication-bus, as it is also typical in other domains
(processor-bus-bottleneck). Given a high-speed communication channel,
the proposed key exchange and rekeying in the $k\!H\!z$-range allows us to use rather weak encryption
algorithms (cf. Section \ref{decision}), as the security may rely on
fast rekeying. Of course, any other more sophisticated encryption
algorithm like AES or 3-DES can also be used. 

The achievable average key-exchange rates of the TPMRAs in the
$k\!H\!z$-range, allow to increase the security through a feasible
frequent key exchange. Short key lifetimes can be realized 
efficiently. Also, any successful online attack must at least
achieve the same performance, requiring significant hardware
expenses. This does not appear to be feasable. Using different keys for encryption and transmission of
different blocks of data, increases the difficulty for an attack on
the encrypted data. 

Due to the small area in the range of one square-millimetre, we regard
the field of application principally as an IP-core in embedded system environments. 
A particular focus can be smartcards or
transponder-based applications such as RFID-systems and devices in
ad-hoc networks \cite{RE00},
in which a small area for cryptographic components is mandatory. 
\section{Summary and Outlook}\label{conc}
We presented a solution for secure communication in embedded system
environments via Tree Parity Machine Rekeying Architectures. 
Our investigations confirm the results as presented in \cite{KKK02}
and stress the advantages of a hardware implementation. 
The silicon area lies within a square-millimetre and
allows to exchange keys of practical size within about a millisecond.
The proposed exchange in parallel to encryption-transmission-decryption also
allows for efficient rekeying schemes and short key lifetimes.   

Next to algorithmic extensions to further increase the security \cite{RKKK02a,RK03,MKKK03,RKSK04}, architectural improvements or variants include a
fully serial realisation with TDMA usage of a single TPM unit. This
further decreases the area consumption but at the cost of an increase
in necessary cycles for one output bit. 
A stream cipher variant, using output bits
directly via the {\em Blum-Blum-Shub} bit generator, 
was suggested already in \cite{KKK02} and
its implementation in hardware is particularly suited for streaming applications.
The relatively small size of the TPMRAs allows an implementation in embedded systems with only small
overhead. They are especially suited for devices of
limited resources and even more in moderate security scenarios. Consequently, the integration of our architectures into such
a system and its practical evaluation is subject to future work.

\bibliographystyle{amsplain}
\bibliography{paper}
\end{document}